\documentstyle[12pt,epsbox,subeqnarray]{article}
\newcommand{\clp}{\clearpage}
\newcommand{\bea}{\begin{subeqnarray}}
\newcommand{\eea}{\end{subeqnarray}}
\newcommand{\be}{\begin{equation}}
\newcommand{\ee}{\end{equation}}

\title{Contraction of a Bundle of Actin Filaments:\\
50 years after Szent-Gyorgyi
\footnotemark[2]}
\author{Ken SEKIMOTO\footnotemark[4] and Hatsumi NAKAZAWA\footnotemark[5]\\
{\it Yukawa Institute for Theoretical Physics,} \\
{\it Kyoto University,Kyoto, 606-01 Japan}\\
}
\date{\null}

\begin{document}
\maketitle
\footnotetext[2]{Invited talk presented at APCTP Inauguration Conference,
June 4-10, 1996, Seoul, Korea}
\footnotetext[4]{\small To whom the correspondence should be addressed;
sekimoto@yukawa.kyoto-u.ac.jp}
\footnotetext[5]{\small Present address: 
International Institute for Advanced Research,
Central Research Laboratories, 
Matsushita Electric Industrial Co., Ltd.}

\begin{abstract}
{\sf
Biological systems are among the most challenging subjects
for theoretical physicists, as well as experimentalists or simulationists.
Physical principles should have been both constraints and guide-lines
for the evolution of living systems over billions of years.
One generally aims at clarifying what physical principles, possibly new ones,
are behind the phenomena of biological interest and at understanding how 
they work within the entire biological world.
In the present talk we describe an example of such an effort. 

Since the discovery of `superprecipitation' by Szent-Gyorgyi's group 
in 1940's,
it has been a long puzzle how an assemblage of actin filaments with random
orientation can contract in the presence of the two-headed myosin molecules 
undergoing actin-activated ATP-hydrolysis reaction. It is widely accepted that
during the contraction the two-headed myosin mediates the relative sliding 
of two actin filaments whose polarity directions are not parallel but rather
anti-parallel. But this fact solely does not account for the shortening.
We propose a dynamical model which, upon numerical simulation, exhibits
the shortening of an bundle of the actin filaments which are initially
dirstributed randomly both in space along a line and in polarity
direction.
In the course of shortening several clusters of actins appears along the bundle.
The model also shows the sorting of the actin filaments according 
to their polarity in the late stage. 
These findings are
in accordance with the recent experiment by Takiguchi.
}
\end{abstract}

\clp
\baselineskip 20pt
\section{Introduction}
There is much  interest in  biological systems from a 
physicists point of view by several reasons.
First, by looking at 
 the diversity and hierarchy of biological phenomena and at the billions of
years of their evolution, 
it is a challenge to unveil universal phenomena or universal
origins of those systems based on physical principles.
For instance, ATP (adenosine triphosphate) is often described
in biology textbook 
as the energy source [energy donner],
as the substrate of transferase [phosphate donner], and 
as the substrate of allosteric enzyme [regulation factor].
We are, however, tempted to search more unified view of the role of ATP since
it should have appeared on the earth initially bearing a single role.
Secondly, the biological systems are the very representatives of complex 
systems. By regarding them as systems of active elements we are inspired 
with many physical ideas and models.

If we view the subjects of biology which have become also the subjects
of physicists, we find that there are some frameworks elaborately introduced 
so that physicists can develop their idea upon it.
Protein folding is studied based upon the Anfinsen's dogma that 
(most) natural proteins rest in their {equilibrium} folded states.
Neural network, despite the prohibiting complexity in reality, is studied
based upon mathematical realization of Hebb's hypothesis. 
Fluctuating membrane is studied on the basis of elasticity theory including 
entropic or Helfrich interaction,
and molecular evolution is studied as stochastic process. 
Morphogenesis and pattern formation have been studied in the framework
of bifurcation theory, etc.
Protein dynamics can be one of the near future target of physicists, being
stimulated by the recent development and need of nanoscale handling of 
soft materials, though the theoretical framework for it is not yet established.

It is said that the biological processes which appear
to be purely physical phenomenon, such as symmetry breakdown or instability
are even sometimes  coded explicitly on the DNA.
It should be, however, still meaningful to ask ``how did such a biological
process happen to appear and become incorporated into evolutionary process?''
Upon the appearance of a new biological function, it should have been quite 
primitive and unsophisticated, which works in barely efficient way 
or in a poorly organized way.
Exploring the mechanism of such primitive functions should then  
be a subject of physics of biological interest as well \cite{prost}.
As such an example, we present in this paper how the system of 
random assemblage of myosin and actin filaments (both being the constituent 
proteins of our muscle) can exhibit stochastic contraction 
phenomena, which is recently studied experimentally in detail \cite{takiguchi}.
\section{System of many myosins and actin filaments: $\qquad$ A paradox}
\begin{figure}[t]
\begin{minipage}{4.03cm}
   \postscriptbox{3.73cm}{6.66cm}{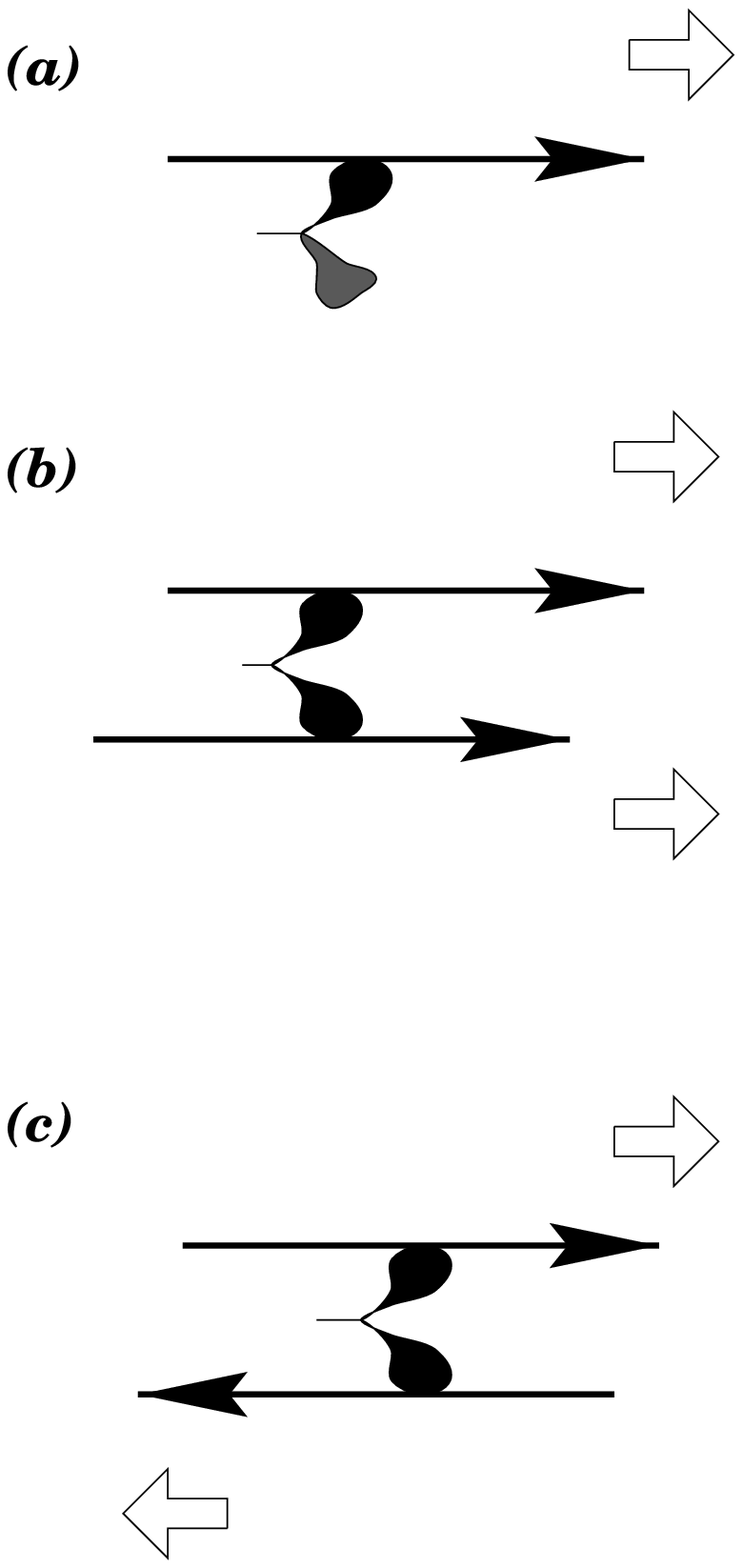}
   \caption{ (a) A head of the two-headed myosin (`seed-leaf')
   translocates an actin filament (thin arrowed line) in the forward 
   direction indicated by an open thick arrow.
   (b) and (c) The action of a myosin onto two filaments.
   \label{base}}
\end{minipage}\hspace{0.5cm}
\begin{minipage}{6.03cm}
   \postscriptbox{5.83cm}{6.66cm}{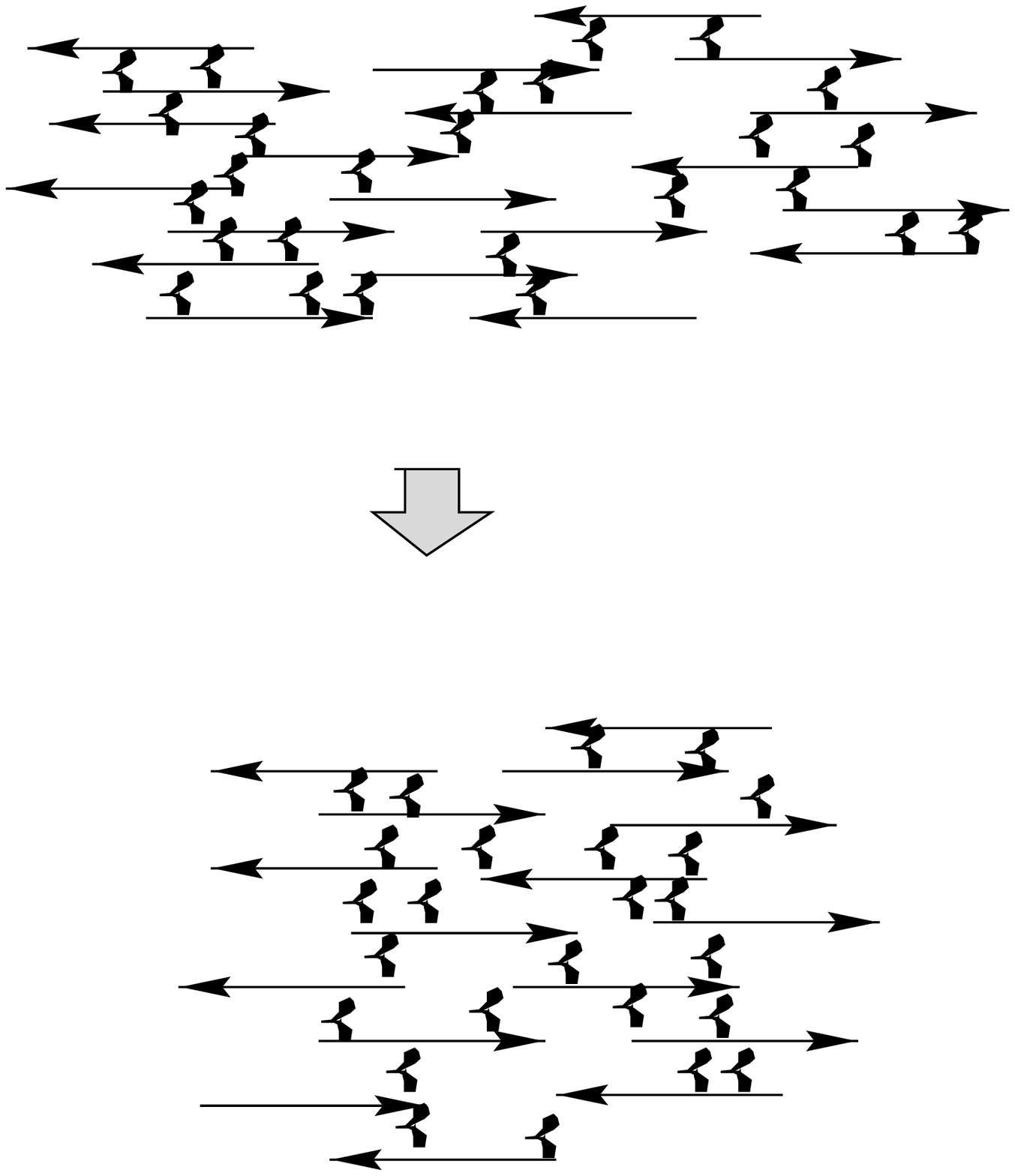}
   \caption{{\it Top:} 
   Uniaxial assemblage, or bundle, of actin filaments, and the myosin
   molecules bound to them.  
   {\it Bottom:} As the myosins translocate the filaments,
   overall shortening of the bundle occurs.
   \label{bundle}}
\end{minipage}\hspace{0.5cm}
\begin{minipage}{4.83cm}
   \postscriptbox{4.83cm}{4.12cm}{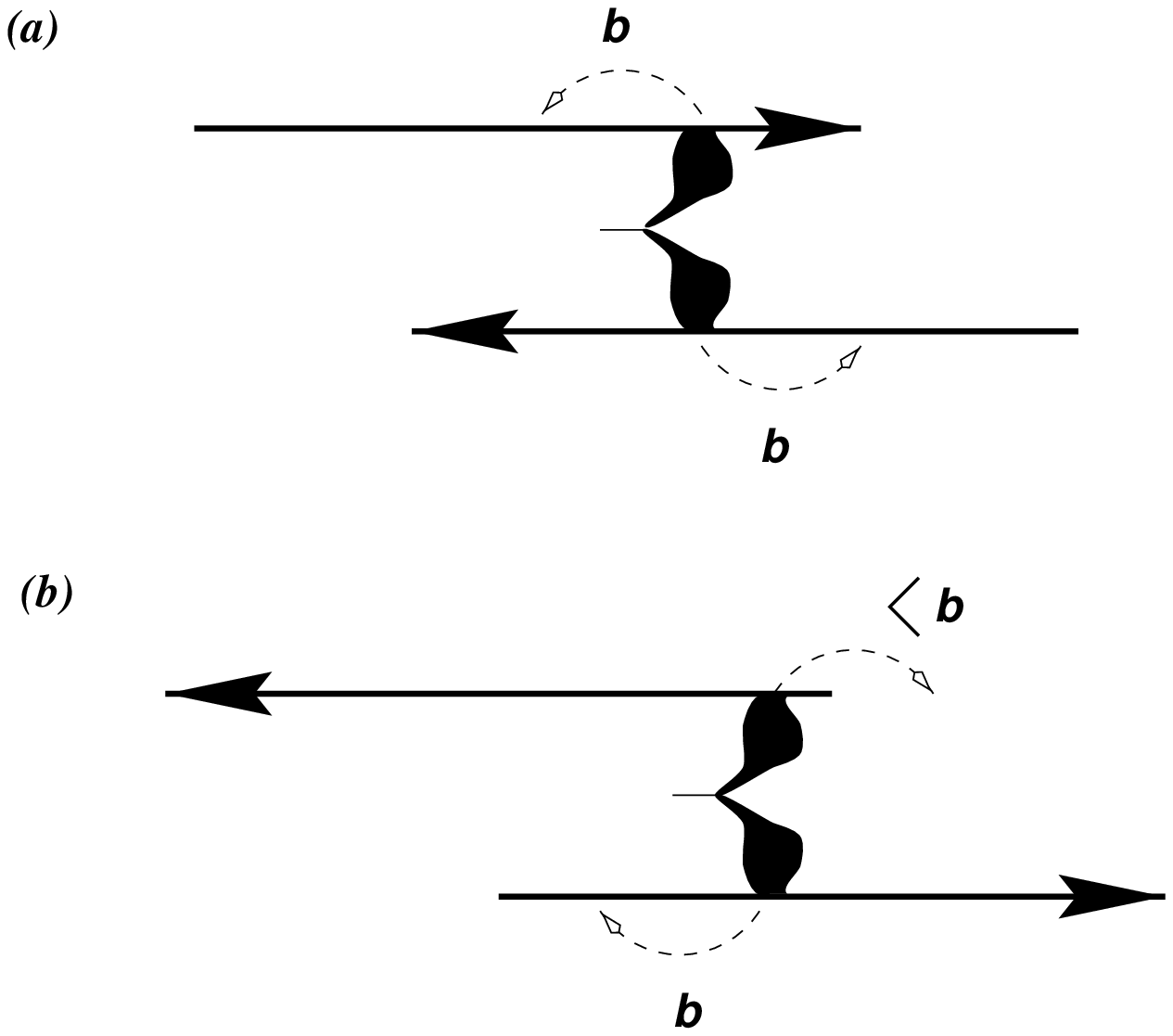}
   \caption{The action of myosin can contributes to either 
   (a) shortening or (b) to elongation of the actin bundle. 
   In the case (b) it can happen that the elongation is interrupted,
   as argued in the text. \label{key}}
\end{minipage}
\end{figure}

We are considering the system consisting of myosins and actin filaments.
Each myosin has two globular heads (shown by the symbols like 
seed-leaves in Fig.~\ref{base}), and each actin filaments
has its own polarity direction (indicated by 
the thin arrow lines in Fig.~\ref{base}).
If a globular head of a myosin is within the reach of an actin filament
in the presence of ATP, the globular head consumes the hydrolysis
energy of ATP to
drive the actin filament to the forward direction (indicated by
the open thick arrow in Fig.~\ref{base}(a)).
The net relative motion of actin filaments is brought by the action of a myosin
only when the myosin is bound to non-parallel pair of filaments 
(Fig.~\ref{base}(c)), but not to parallel pair (Fig.~\ref{base}(b)).

We focus here on the recent experiment by Takiguchi \cite{takiguchi},
He prepared a bundle of many actin filaments which are assembled uniaxially
but randomly with respect both to
their position and to their polarity direction.
It has been demonstrated \cite{takiguchi} that the bundle of actin filaments
undergoes longitudinal shortening in the presence of myosin and ATP molecules.
We will describe the experimental procedure of \cite{takiguchi} in more 
detail.
First a long and thick bundle of many actin filaments is
prepared in methyl cellulose aqueous solution.
To this bundle the two-headed myosins (so called heavy meromyosin, or HMM) 
and an abundance of ATP molecules are added.
The bundle then starts to contract slowly in length, while it thickens so
as to conserve its volume (Fig.~\ref{bundle}).
This shortening often occurs in a wiggling way. 
After the bundle has shortened appreciably, several needle-like subbundles 
appear from the main bundle. 
It has been shown that in these subbundles the polarity of the actin 
filaments is not random but is oriented outward with respect to 
the original bundle.
%

Experiment like this dates back to late 40's, 
when Szent-Gyorgyi's group discovered so called superprecipitation,
the phenomenon that a 
three-dimensional random assembly of actin filaments and myosins 
shrinks dramatically after the addition of a certain amount of ATP \cite{SG}.
Such experiment has been recently also done and refined \cite{fujime}.
Takiguchi's setup \cite{takiguchi} may be regarded as a more idealized one to see
how the contraction occurs.
Such an ideal random distribution is realized only in {\it in vitro} 
experiment, but the situations more or less like
this have been found in nature such as in the contractile rings that
appear during the mitotic period of cell division cycle \cite{mabuchi}
or in stress fibers observed in the locomoting cells during their 
contraction period \cite{mitchison,SS}.
The experiments mentioned above, therefore, could be regarded as
a simulation of {\it in vivo} systems or, at least, as
a hypothetical simulation of evolutionally primitive stages of muscle
contraction or cell motility.
The question how this primitive system undergoes shortening has, however,
not been studied for a long time since the discovery by Szent-Gyorgyi's group.
It is because the highly organized structure of muscle 
was found \cite{HH} shortly after the former discovery, 
and the main stream of muscle study has been focused 
towards a dynamics of single globular head of myosin and its regulation 
mechanism \cite{cell}.

It is Hatano who seriously questioned how the actin bundle can shorten
in the primitive situation like in Fig.~\ref{bundle}, and he came across the  
following paradox \cite{hatano}:
When the sliding of the oppositely oriented actin filaments 
starts from the state shown in Fig.~\ref{key}(a), 
the overlap between the two filaments would increase, leading
to the shortening of the bundle. On the other hand, when the sliding of the
filaments starts from
the state shown in Fig.~\ref{key}(b), the action of the myosin 
would decrease the overlap between the filaments, leading to the elongation 
of the bundle. 
Since the both situations should occur equally likely in a 
bundle, there would be no net shortening at all.
In fact so-called bipolar kinesin, the other motor protein closely related
to myosin, is discovered to appear during the cell division process 
and this protein is thought to act to {\it separate} the two spindle-poles
by the mechanism shown in Fig.~\ref{key}(b) \cite{bipolar}.
We would note that the above paradox cannot be lifted by considering 
the effect of simultaneous action of many myosins to 
an actin filament, as it occurs experimentally 
as far as we assume the {\it continuous} action of myosin molecules on the actin 
filaments, while such model could predict the undulational instability 
of filament density \cite{HN}.
\section{Simple model and simulation}
Our simple idea to resolve Hatano's paradox is to take into account 
the finite distance, say $b$, by which a globular head of myosin can {\it
continuously} drive a single actin filament (Fig.~\ref{key}). 
The limitation of this distance may come from the dynamic fluctuation of the myosin heads
as well as by the fluctuation of the lateral arrangement of actin 
filaments within a bundle. Our reasoning for the shortening is as follows:
If the myosin acts in the situation of Fig.~\ref{key}(a)
each globular head can translocate the respective filament fully by the
distance $b$ on the average
(hereafter we assume that $b$ is sufficiently smaller than the length of each
filament, which we denote by $\ell$), while in the situation of Fig.~\ref{key}(b)
the translocation of actions by myosin can be interrupted when one of its two 
heads meets with the rear end of an actin filament.
The elongation of the bundle to which the filaments shown in Fig.~\ref{key}(b)
belong is, therefore, less extensive than the shrinking of the bundle to 
which the filaments shown in Fig.~\ref{key}(a) belong.
The interruption of the elongation will occur by the probability
proportional to $b/\ell$ in the approximation 
up to the lowest order of $b/\ell\ll 1$. 
The net shrinkage per each action of myosin will then be roughly scaled by 
$\sim b\cdot b/\ell$ if a single myosin acts to the pair of filaments.
Actually the shortening by this mechanism should be 
still less efficient due to the presence of other myosins interacting 
with those actin filaments.
We believe, however, that the basic mechanism of shortening may be captured 
by the present simple model.
\begin{figure}[t]
\begin{minipage}{7.5cm}
   \postscriptbox{7.5cm}{10cm}{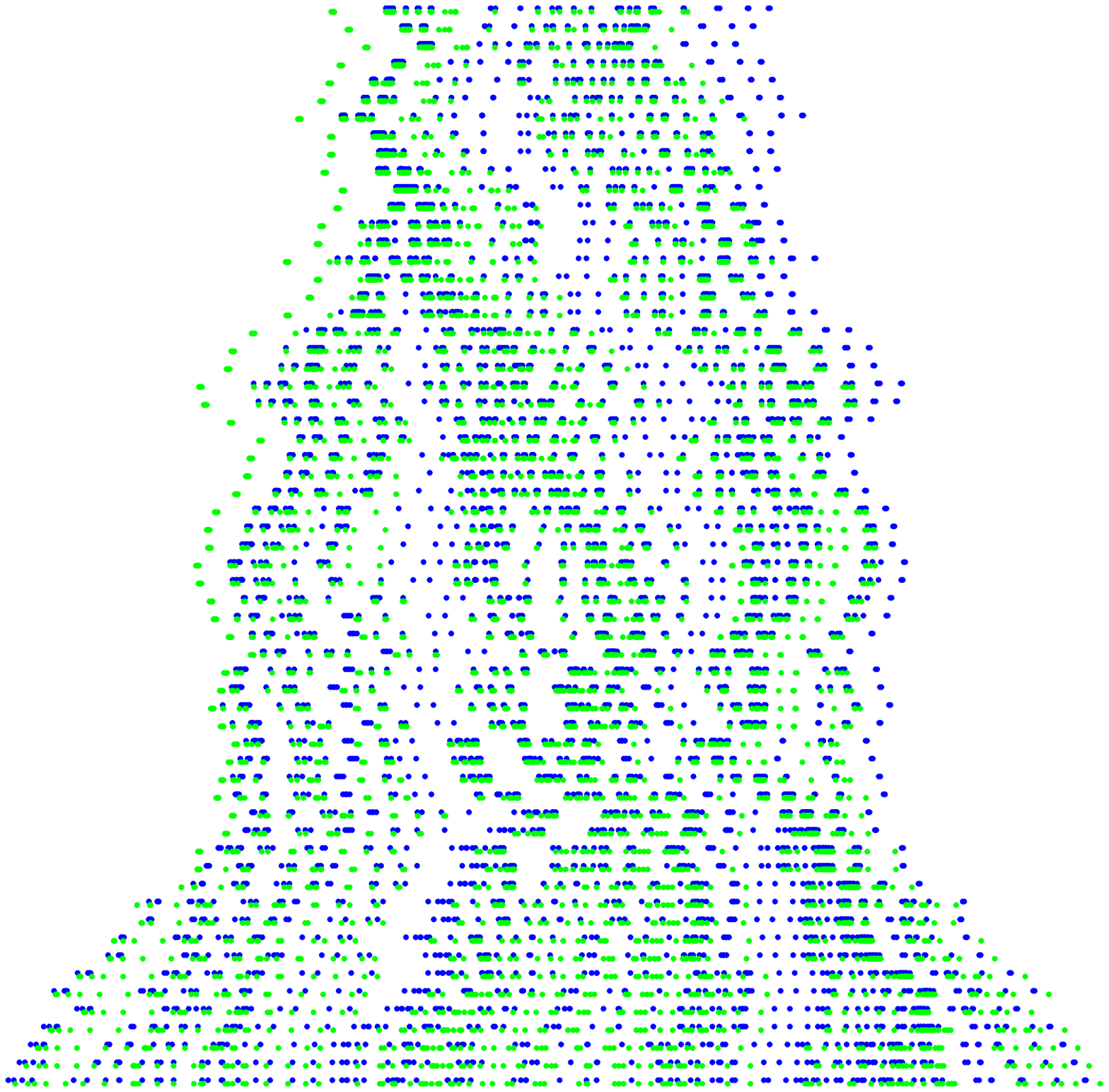}
   \caption{Snapshots of actin distribution; the blue dots indicate the
   center position of actins oriented rightward, and the green ones 
   indicate those oriented leftward. 
   Time proceeds from the bottom to top. 
   For the parameter values used, see the text. \label{demo34}}
\end{minipage}\hspace{0.5cm}
\begin{minipage}{7.25cm}
   \postscriptbox{7.25cm}{10cm}{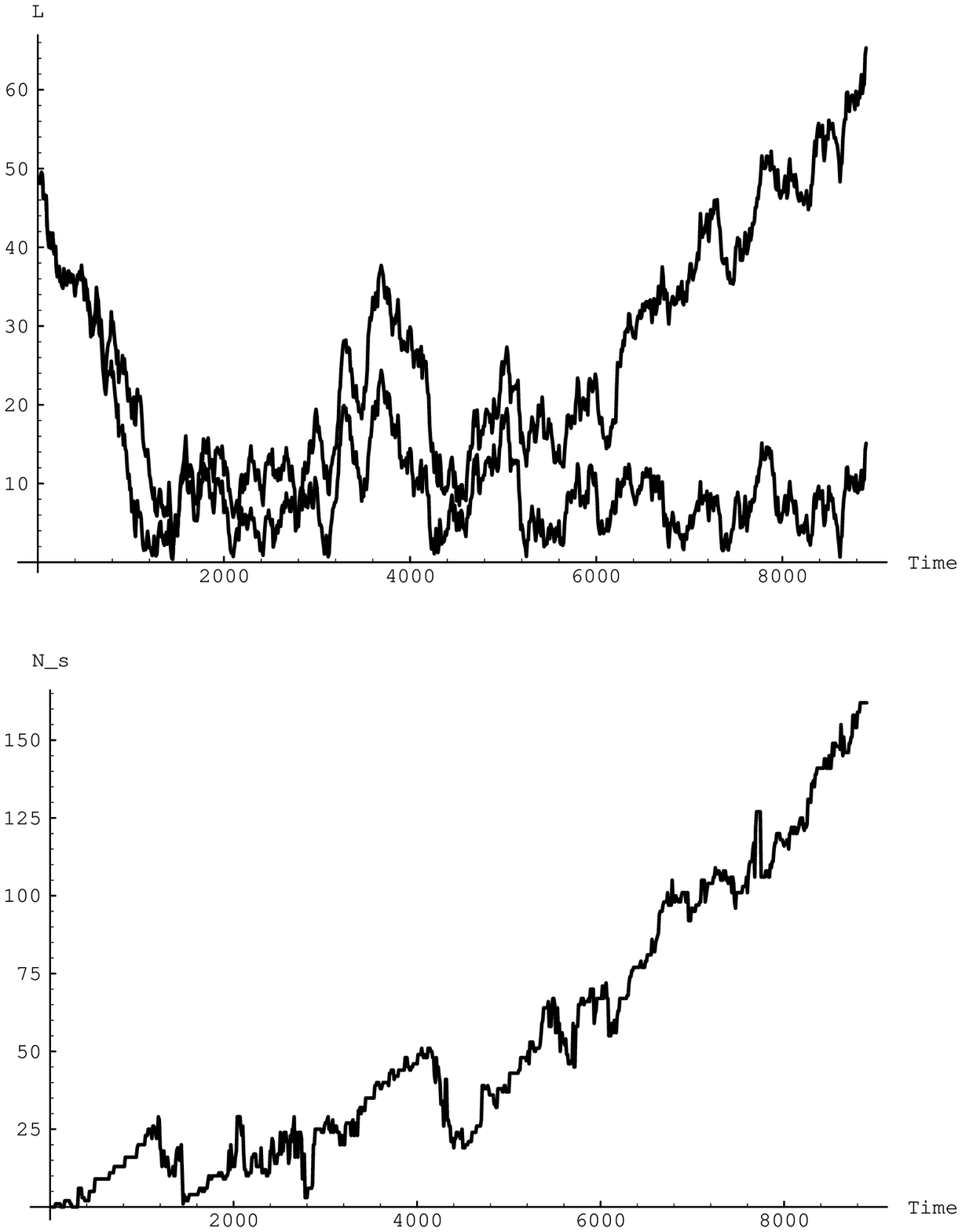}
   \caption{ {\it Top:} Evolution of the length of whole bundle 
   including the polar arms (the upper curve) and the length of the bundle 
   less the polar arms (the lower curve). {\it Bottom:} The evolution of
   the number of actin filaments sorted out into the polar arms.
   The total number of actins, $N$, is 1000 in this calculation.
   \label{hys31}}
\end{minipage}
\end{figure}
\begin{figure}[b]
\begin{minipage}{7.25cm}
   \postscriptbox{7.25cm}{12cm}{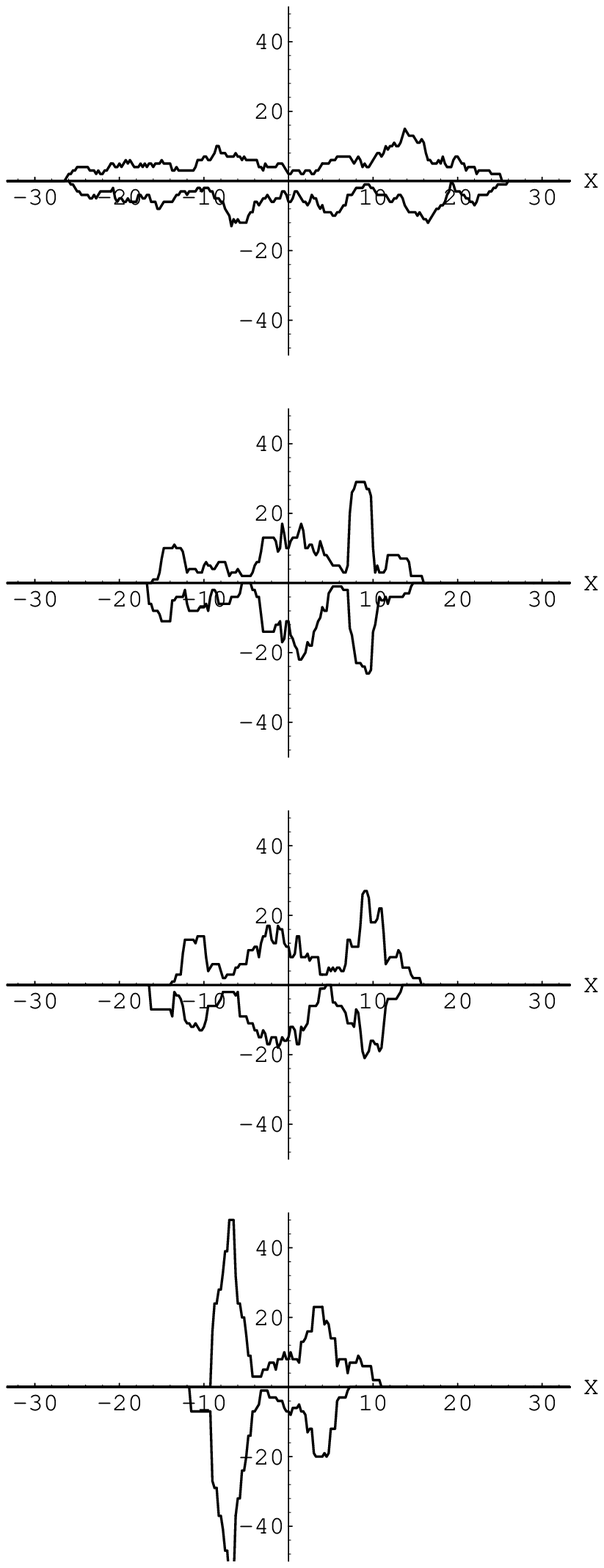}
   \caption{ The snapshots of density profiles of the actin filaments
   oriented rightward (upper curve) and those oriented leftward (lower curve),
   taken from the data shown in Fig.~\protect\ref{demo34}.
   The time lapse is such that the average times of the unitary actions 
   undergone by each filament are, respectively, 0, 1, 2, and 3 from 
   the top to the bottom.
    \label{snap}}
\end{minipage}\hspace{0.5cm}
\begin{minipage}{7.25cm}
   \postscriptbox{7.25cm}{9cm}{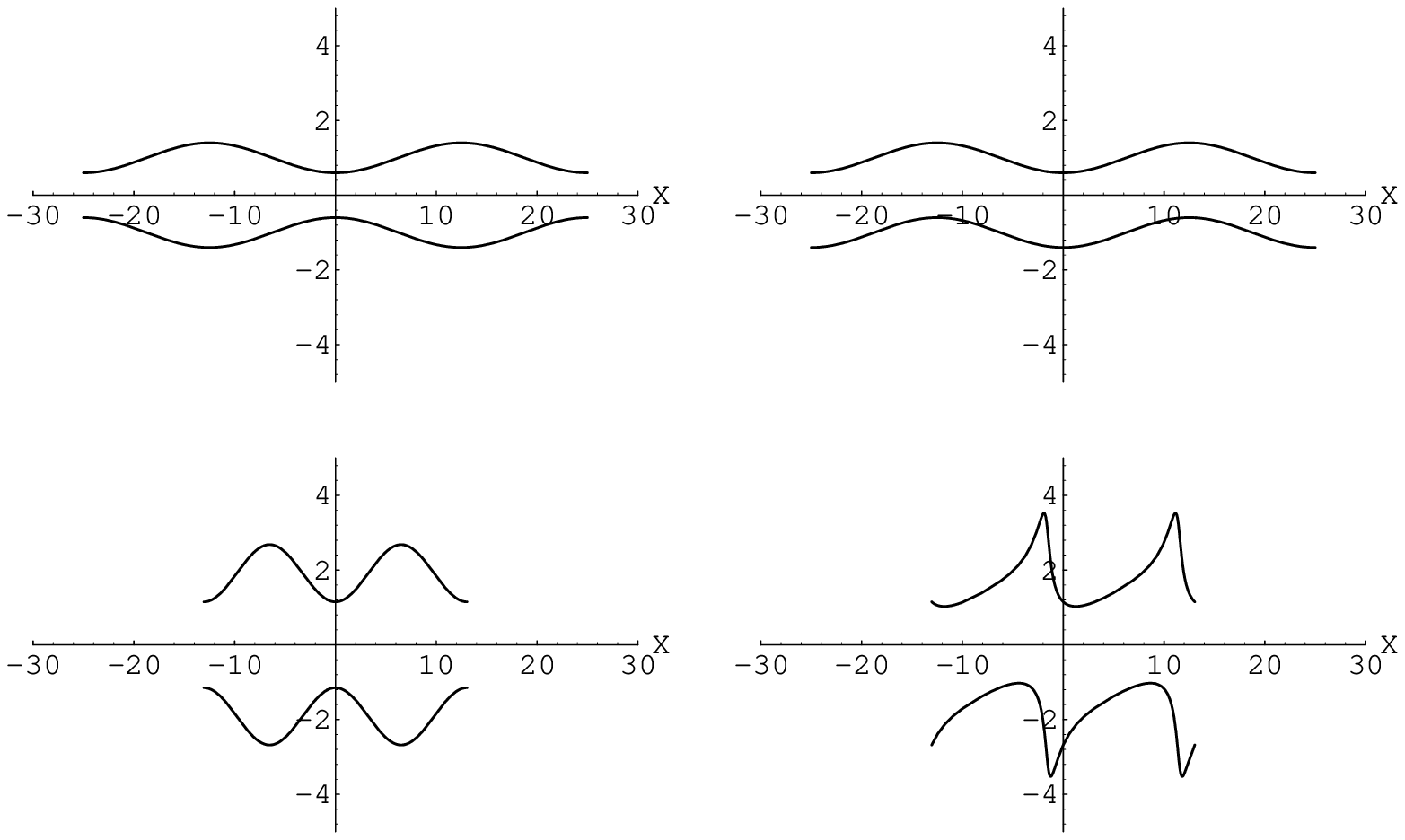}
   \caption{ Solutions of the evolution equation 
   (\protect\ref{eq:pde}) in the text, from the two initial conditions
   (the top raw). The time lapse of the evolved states (the bottom raw)
   are the same for both cases.
    \label{pde}}
\end{minipage}
\end{figure}

We performed a numerical simulation based on the simple idea
described in Fig.~\ref{key}. The algorithm of the simulation is as follows:
First we distribute randomly $N$ actin filaments of the 
length $\ell$ over an interval $-\frac{L_0}{2}\leq x \leq \frac{L_0}{2}$  
along the $x-$axis. 
We choose the parameters so that $N\ell\gg L_0$ holds 
to assure substantial overlapping of the filaments along the $x$-axis.
For evolution, we define the `unitary action' by a two-headed myosin: 
{\it (1)} We chose randomly a spatial point,
say at $x=x_M$, where the myosin translocates a pair of actins (see (2) below). 
{\it (2)} Among all the filaments that extend through the point $x=x_M$
we then choose randomly a pair of anti-parallel actin filaments.
One of the chosen filament is oriented toward the positive $x$ direction 
(i.e., rightward) and is centered at, say, $x_+$,
while the another chosen filament is oriented toward the negative $x$ 
direction (i.e., leftward) and is centered at, say, $x_-$. 
{\it (3)} We move these two filaments by the same distance but in the opposite 
direction, according to the scheme described in Fig.~\ref{key}. 
As seen from this figure the distance of sliding is given by
$U(x_+,x_-,x_{\rm M})\equiv$ ${\rm Min}[b,$ $x_{\rm M}-x_+ 
+{\ell}/{2},$ $x_- +{\ell}/{2}-x_{\rm M}]$.
{\it (4)} As for the rest of the filaments in the bundle, those filaments 
in the region of $x>{x_c}(x_+,x_-) \equiv (x_+ +x_-)/2$ are displaced 
by $+U$ [$-U$] if
$x_+>x_-$ [$x_+<x_-$], respectively. Also, those filaments in the region of 
$x<{x_c}(x_+,x_-)$ are displaced inversely so that $U$ in the last sentence is 
replaced by $-U$.
The evolution of the system is obtained by applying this unitary action
from (1) to (4) repeatedly.

The result of the simulation is represented by the distribution of 
filaments position (Figs.~\ref{demo34}) and by the density of filaments 
along $x$-axis (Fig.~\ref{snap}). 
The parameters used there are $N=200$, $\ell=0.25$, and $L_0=50$.
These values  are comparable to 
the experimental values (in unit of $\mu m$ for the lengths) \cite{takiguchi}. 
The total time lapse of the evolution is such that each filament undergoes,
on the average, six times the unitary action of myosin.
These figures reveal the overall shortening of the assembly of filaments, 
and also shows the clustering of the filaments in rather symmetric fashion with
respect to their polarity.
In Fig.~\ref{hys31} we show another run with extended time lapse.
There appear the `arms'  from both ends of the bundle,
which consist of filaments with unique outward polarities. 
The length of these arms and the number of the 
filaments in these arms increase in time.
Although the simulation is restricted in one dimensional space, 
the characteristic features of the evolution thus found are in agreement 
with experimental observation of (i) shortening of the main bundle,
(ii) inhomogenization of its thickness and (iii) generation of polar subbundles.
\cite{takiguchi}.
\section{Construction of continuum model}
The above algorithm of simulation can be cast into the evolution equation 
for the densities of actin filaments, $\rho_+(x_+)$ and  $\rho_-(x_-)$,
oriented rightward and leftward, respectively.
For this we needed to assume the smallness of the ratio, $b/\ell\ll 1$, and
restrict ourselves to the limit of weak spatial heterogeneity, 
$|\rho^\prime_\pm|\ll \ell \rho_\pm$. 
Suppose that a myosin is at $x=x_{\rm M}$ and starts to exert the unitary action
to a pair of anti-parallel filaments.
Let us denote by ${\cal P}(x_+,x_-;x_{\rm M})dx_+ dx_-$ the probability 
that these two filaments are centered  
at $x=x_+\sim x_+ +dx_+$ (the rightward filament) and at 
$x=x_-\sim x_- +dx_-$ (the leftward filament), respectively.
In the mean field approximation, this is given as
\be
{\cal P}(x_+,x_-;x_{\rm M})=\frac{
\rho_+(x_+) \rho_-(x_-)
 \theta(\frac{\ell}{2}-|x_+ -x_{\rm M}|)
 \theta(\frac{\ell}{2}-|x_- -x_{\rm M}|)
}{
\int_{|x_+^\prime -x_{\rm M}|<\frac{\ell}{2}}dx_+^\prime
\int_{|x_-^\prime -x_{\rm M}|<\frac{\ell}{2}}dx_-^\prime
\rho_+(x_+^\prime) \rho_-(x_-^\prime)
},
\ee
%
\noindent
where we introduced a step function $\theta(z)=1$  for  $z>0$ and $\theta(z)=0$
for  $z\leq 0$.
Using this distribution we introduce the weighed average 
of any function, say ${\cal O}(x_+,x_-,x_{\rm M})$,
over $x_+$ and $x_-$ 
with  $x_{\rm M}$ being fixed, ${<{\cal O}>}_{x_{\rm M}}$, as 
${<{\cal O}>}_{x_{\rm M}}\equiv$ 
$\int dx_+ \int dx_- {\cal P}(x_+,x_-;x_{\rm M})$ 
${\cal O}(x_+,x_-,x_{\rm M})$.
Then the displacement field $u(x;x_{\rm M})$ of the actin densities 
caused by a unitary action of myosin at $x_{\rm M}$ is
given by
$u(x,x_{\rm M})=$  $<U(x_+,x_-,x_{\rm M})$  ${\rm sgn}(x_+-x_-)$
 ${\rm sgn}(x-{x_c}(x_+,x_-)){>}_{x_{\rm M}}$,
where ${\rm sgn}(z)\equiv 2\theta(z)-1$.
Using the gradient expansion of the densities of actin filaments, 
$\rho_+(x_{\rm M}+X_+)\rho_-(x_{\rm M}+X_-)\simeq$
$\rho_+(x_{\rm M})\rho_-(x_{\rm M})+$ 
$\rho_+^\prime(x_{\rm M})\rho_-(x_{\rm M}) X_+ +$  
$\rho_+(x_{\rm M})\rho_-^\prime(x_{\rm M}) X_- +$~.~.~.
(the prime denotes to
take the spatial derivative), 
the weighed average can be evaluated up to  the first order
approximation to give
\be
u(x,x_{\rm M})  \simeq 
    \left[ -\frac{b^2}{\ell} + \frac{b \ell}{6}\left( 
    \frac{\rho_+^\prime(x_{\rm M})}{\rho_+(x_{\rm M})} 
    -\frac{\rho_-^\prime(x_{\rm M})}{\rho_-(x_{\rm M})}  \right)\right]
    {\rm sgn}(x-x_{\rm M}).
\ee
Here we have noted that ${\rm sgn}(x-{x_c}(x_+,x_-))$ can be safely replaced
by ${\rm sgn}(x-x_{\rm M})$ in the coarse grained 
description which deals with only the length scales larger than $\ell$.
The zeroth order term $-b^2/\ell$ in the angular bracket represents 
the tendency of shortening described already, and the second term
represents the correction due to the spatial inhomogeneity of the
filament densities. 
If, for example, $\rho^\prime_+<0$ and 
$\rho^\prime_->0$ hold at $x=x_{\rm M}$, the latter term predicts
that the shortening is enhanced compared with the homogeneous case. 
It is understandable since $\rho^\prime_+<0$ and $\rho^\prime_->0$
imply the situations like 
Fig.~\ref{key}(a) is be more likely to be found at $x_{\rm M}$ than those like 
Fig.~\ref{key}(b).

The mean drift velocity of the bundle, ${\bar v}(x)$, is obtained as the
integration of $u(x,x_{\rm M})$ with respect to $x_{\rm M}$ 
multiplied by the frequency factor $\kappa(x_{\rm M})$ with which
myosins exert the unitary actions per unit time and per unit interval 
along the $x-$axis.
The evolution equation for $\rho_\sigma$ ($\sigma=\pm$) is then
$\frac{\partial}{\partial t} \rho_{\sigma}=$
$-\frac{\partial}{\partial x} \left[ {\bar v}(x) \rho_{\sigma} \right]$.
Hereafter we the simplest choice that the factor $\kappa(x_{\rm M})$ 
is an overall constant, say $\kappa(x_{\rm M})=\kappa_0$.
This case is that one can solve  the evolution equation analytically and,
at the same time, that the essential feature of shortening and clustering of
the bundle is preserved (see below).
Performing the integration with respect to $x_{\rm M}$ 
the evolution equation becomes;
\be
\frac{\partial}{\partial t}\rho_{\sigma}(x,t)=
 - \frac{\partial}{\partial x}\left\{ \kappa_0 
\left( -\frac{2 b^2}{\ell} x + \frac{b \ell}{3} \log \left[  
\frac{\rho_+(x,t)}{\rho_-(x,t)}
\right]\right) 
\rho_{\sigma}(x,t)  \right\}, \qquad \sigma=\pm.
\label{eq:pde} \ee

The solution of initial value problem can be
given via parametric representation as follows:
 \be 
  {\hat x}(X,t)=X e^{-\frac{2 \kappa_0 b^2}{\ell} t}
 + \frac{\ell^2}{6 b}\left( 1- e^{-\frac{2 \kappa_0 b^2}{\ell} t} \right)
  \log \left[\frac{\rho_+ (X,0)}{\rho_- (X,0)}\right],
  \ee
  \be
  \rho_{\sigma}({\hat x}(X,t),t)=\rho_{\sigma}(X,0) 
  {\left[ \frac{\partial {\hat x}(X,t)}{\partial X}  \right]}^{-1}, 
  \qquad \sigma=\pm.
  \ee
Figure \ref{pde} shows two examples of the solution of  
(\ref{eq:pde}), the one starting from the actin densities 
with  in-phase undulation (the left column) 
and the other one starting with 
anti-phase undulation (the right column), respectively.
If we neglected the logarithmic correction term in (\ref{eq:pde}), the solution would
simply represent the affine contraction of the bundle, i.e.,
$ \rho_{\sigma}(x,t)= $
$\frac{t_c}{t_c-t}  \rho_{\sigma} \left(\frac{t_c}{t_c-t}x,0 \right)$ 
for $\sigma=\pm$,
where $t_c={(2 \kappa0 b^2/\ell)}^{-1} L_0$ is the time 
at which the bundle with the initial length of $L_0$ shrinks down to 
a point.
As seen from Fig.~\ref{pde} the correction term acts to promote the 
clustering of actin filaments of both rightward and leftward polarity.
Two remarks are in order here: 
We should note that the simulation described in the previous section, 
as well as the experiments with low myosin concentration, 
would correspond to the slightly different choice of $\kappa(x_{\rm M})$,
that is, $\kappa(x_{\rm M})=$ $\kappa_0 L_0/L(t)$, where $L(t)$ is the total 
length of the bundle at time $t$.
This overall factor, ${L(t)}^{-1}$
would change the time scale of the evolution of Fig.~\ref{pde},
but does not change the features of the evolution of the density profiles.
We would also note that the generation of the arm cannot be handled within 
the present approximation in which the spatial variation of 
actin densities is assumed to be small. 
 
\section{Summary}
We have proposed a simple model for the of contraction
of the random uniaxial assembly of actin filaments, mediated by the two headed
myosin molecules which translocates anti-parallel actin filaments.
Simulation result have agreed at least in the qualitative level with 
the experimental observation: the shortening of the actin bundle, 
the clustering of density and also the generation of polar arms.
The experimental situation studied here may well correspond
to the stage of evolution where the collective transport by motor proteins
had first come into existence in the biological world.
More generally, it would be interesting to study from physicists' viewpoint
how a function, in its most primitive form, has been first acquired 
by biological systems at any level of evolutionary history; 
the problem how an allosteric enzyme have acquired the
function to translocate the other molecule is a challenging problem 
in this respect.
\section*{Acknowledgements}
The authors gratefully appreciate K. Takiguchi for the kind introduction to his 
experiments.
They also thank very much F. Oosawa, Y. Oono, K. Tawada and M. Ishigami 
for valuable critical comments on the subject.
Lastly but not least one of the author (K.S.)  would like to acknowledge the 
organizers of the Inauguration Conference of APCTP for the enjoyable meeting and 
their hospitality.

\end{document}